\documentclass[12pt,a4paper]{article}

\usepackage[T2A]{fontenc}
\usepackage[cp1251]{inputenc}
\usepackage{authblk}
\usepackage[a4paper,margin = 0.62in]{geometry}

\usepackage{setspace}
\usepackage{graphicx}
\usepackage{amsmath,amssymb,bbm}
\usepackage{color}

\usepackage[toc,page]{appendix}

\def\beq{\begin{equation}}
\def\eeq{\end{equation}}
\def\bea{\begin{eqnarray}}
\def\eea{\end{eqnarray}}

\def\MM{{\mathfrak M}}

\def\RR{{\mathcal{R}}}

\title{Noether's theorem in non-local field theories}

\author{M. I. Krivoruchenko$^{1,2,3}$ and A. A. Tursunov$^{1,4}$}
\affil{
$^1$Bogoliubov Laboratory of Theoretical Physics\\  Joint Institute for Nuclear Research,
141980 Dubna, Russia\\
$^2$Institute for Theoretical and Experimental Physics\\
B. Cheremushkinskaya 25, 117218 Moscow, Russia\\
$^3$Moscow Institute of Physics and Technology, 141700 Dolgoprudny, Russia \\
$^4$Institute of Physics, 
Silesian University in Opava, \\ Bezru{\v c}ovo n{\'a}m.~13, CZ-74601 Opava, Czech Republic\\
}

\date{}

\begin{document}

\maketitle

\abstract{
Explicit expressions are constructed for
a locally conserved vector current associated with a continuous internal symmetry
and
for energy-momentum and angular-momentum density tensors associated with the Poincar\'e group
in field theories with higher-order derivatives and in non-local field theories.
An example of non-local charged scalar field equations with broken C and CPT symmetries is considered.
For this case, we find simple analytical expressions for the conserved currents.
}



\section{Introduction}

\renewcommand{\theequation}{1.\arabic{equation}}
\setcounter{equation}{0}
$\;$ \vspace{-12pt}

According to Noether's theorem \cite{Noether}, the invariance of the Lagrangian of a physical system%
with respect to continuous transformations leads to the existence of conserved charges.
In its standard form, Noether's theorem refers to local field theories with derivatives of no higher than second order in the field equations.

Quantum field theories
with higher derivatives are used for intermediate regularization procedures (see, e.g., \cite{Faddeev1993}).
The low-energy regime of QCD is known to be successfully described by chiral perturbation theory
based on a power-series expansion in derivatives \cite{Gasser1985,Scherer2003}.
Field theories with higher derivatives are also discussed in the context of general relativity \cite{Szabados2009}.

The CPT theorem tells that the CPT symmetry violation can be related to non-local interactions.
Low-energy nuclear and atomic experiments provide strict constraints on the scale of a possible violation of CPT symmetry.
A simple classification of
the effects of the violation of the C, P and T symmetries and their 
combinations is presented by Okun~\cite{Okun:ARXIV:2002}.
A class of inflationary models is based on a non-local field theory \cite{Barnaby2008}.

In this paper, the question of whether one can generalize Noether's theorem to non-local field theory is discussed.

As an initial step, we consider a Lagrangian that contains, along with a field $\Psi = (\phi,\phi ^{\ast})$,%
its higher derivatives $\partial _{\mu _{1}}\ldots \partial _{\mu _{n}}\Psi $ up to order $n \geq 1$:
\begin{equation}
\mathcal{L}=\mathcal{L}(\Psi ,\partial _{\mu _{1}} \Psi,\ldots ,\partial _{\mu _{1}}\ldots \partial _{\mu _{n}} \Psi).  \label{lag-loc}
\end{equation}%
The Lagrangian given in (\ref{lag-loc}) is still local because it is a function of the
field and its finite-order derivatives evaluated at a single point in
space-time. To obtain a non-local field theory, one must include
in (\ref{lag-loc}) a dependence on an infinite number of field derivatives,
i.e., by considering the limit $n\rightarrow \infty $.

In the remainder of this paper, we use a system of units such that $\hbar =c =1$.
Indices $\mu ,\;\nu ,...$, denoted by Greek letters from the middle of the alphabet, run from $0$ to $3$.
Indices $\alpha ,\;\beta ,...$ denote the spatial components of tensors and run from $1$ to $3$.
We use a time-like metric $g_{\mu \nu }= \mathrm{diag}(+1,-1,-1,-1)$, and
indices enumerating members of internal-symmetry multiplets are suppressed.


\section{Symmetries and the conserved currents}

\renewcommand{\theequation}{2.\arabic{equation}}
\setcounter{equation}{0}
$\;$ \vspace{-12pt}

All observable quantities can be expressed in terms of fields and
their combinations. The fields that appear in the Lagrangian, in general, belong to
a representation space of the internal symmetry group.
Linear transformations of fields related to the internal symmetry group do not
affect physical quantities, which is the case considered in the present
paper. Thus, for infinitesimal transformations related to internal%
symmetries, one can write the transformation matrix as follows
\begin{equation}
U(\omega )=1-i\omega ^{a}T^{a},  \label{int-sym}
\end{equation}%
where the $\omega ^{a}$ are a set of infinitesimal real parameters and
the $T^{a}$ are generators of group transformations. If the matrix $U$ is unitary, then the $T^{a}$ are Hermitian
matrices. For the $U(1)$ symmetry group, $T^{a}=1$, and for $SU(2)$, the $T^{a}$ are the Pauli matrices.


In the general case, along with the internal symmetry of a system, one
must consider the existence of external symmetries related
to the invariance of physical quantities with respect to
translations and the Lorentz transformations.
Invariance under space-time translations leads to
energy-momentum conservation, whereas Lorentz invariance gives rise to the
conservation of angular momentum. For an infinitesimal element of the Lorentz group,
coordinate transformations can be realized by means of the matrix
$a_{\nu}^{\mu }=\delta _{\nu }^{\mu }+\varepsilon _{\cdot \nu }^{\mu \cdot }$,
where $\varepsilon ^{\mu \nu }$ is an infinitesimal antisymmetric tensor.
This implies that the infinitesimal Lorentz transformation matrix in the representation space of the field
can be written in the most general form as follows:
\begin{equation}
S(a) = 1 - \frac{i}{2} \varepsilon_{\mu \nu } \Sigma ^{\mu \nu },
\end{equation}%
where $\Sigma^{\mu \nu }$ is a matrix defined by the transformation properties of the field.
Thus, the complete transformations of the coordinates and the field
corresponding to the internal and external symmetries can be expressed in
matrix notation as follows:
\begin{eqnarray}
x^{\prime } &=& ax+b, \label{x-transform} \\
\Psi ^{\prime }(x^{\prime }) &=& U(\omega )S(a)\Psi(x),  \label{Psi-transform}
\end{eqnarray}
where the notation used in Eq.~(\ref{x-transform}) dictates a particular order of the transformations, namely, translation is
performed after Lorentz transformation. In the opposite case, one must
use $x^{\prime }=a(x+b)$. In the particular case of a scalar field,
we obtain the simple expression $\phi ^{\prime }(x^{\prime}) = \phi (x)$,
where $\phi $ is a scalar with respect to the internal group
of symmetries and with respect to the Lorentz transformations.
The field $\Psi (x)$ in general belongs to a nontrivial representation
of the internal symmetry group and a Poincar\'e group representation.

For the infinitesimal parameters $\omega ^{a} $, $\varepsilon ^{\mu \nu}$
and $b^{\mu }$, the variation of the field takes the form
\begin{equation} \label{delta-Psi}
\begin{split}
\delta \Psi (x) &= \Psi ^{\prime }(x)-\Psi (x)  \\
&= S(a)U(\omega )\Psi (a^{-1}x-b)-\Psi (x)   \\
&= \left( -i\omega ^{a}T^{a}\right) \Psi (x)-b^{\mu }\partial _{\mu }\Psi
(x) - \frac{i}{2}\varepsilon ^{\mu \nu } \left(\Lambda_{\mu \nu } + \Sigma _{\mu \nu }\right) \Psi (x).
\end{split}
\end{equation}
The intrinsic symmetry generates variation $\delta \Psi =(\delta \phi, ~\delta
\phi ^{\ast })$ with $\delta \phi =-i\omega ^{a}t^{a}\phi $ and $\delta
\phi ^{\ast }=i\omega ^{a}\tilde{t}^{a}\phi ^{\ast }$. We thus use $%
T^{a}=(t^{a},-\tilde{t}^{a})$.
The spin generators act as $\delta \phi =-\frac{i}{2}\varepsilon _{\mu \nu }\sigma ^{\mu \nu }\phi $
and                        $\delta \phi^{\ast }=\frac{i}{2}\varepsilon _{\mu \nu }\sigma ^{\mu \nu \ast }\phi^{\ast }$.
The rotation operators are defined by
$\Lambda_{\mu \nu } = (\mathcal{R}_{\mu \nu },-\mathcal{R}_{\mu \nu })$ where $\mathcal{R}_{\mu \nu } = x_{\mu}i\partial _{\nu }-x_{\nu }i\partial _{\mu }$ and
$\Sigma ^{\mu \nu }=(\sigma ^{\mu \nu},-\sigma ^{\mu \nu \ast })$.
The order of the transformations is as follows: the matrix $U(\omega )$ applies first,
followed by the Lorentz transformations and then translation.
However, the order of the matrices $S(a)$ and $U(\omega )$ is interchangeable because the transformations of internal symmetries commute with those of
external symmetries. Thus, the first term on the right-hand
side of Eq.~(\ref{delta-Psi}) corresponds to transformations of internal
symmetries, the second corresponds to translations, and the third corresponds to the Lorentz
transformations.

Returning to Eq.~(\ref{lag-loc}) for the infinitesimal parameters $\omega
^{a} $, $\varepsilon ^{\mu \nu }$ and $b^{\mu }$, one can now write the
variation of the Lagrangian, $\delta \mathrm{{\mathcal{L}}}(x)=\mathrm{{%
\mathcal{L}}}^{\prime }(x)-\mathrm{{\mathcal{L}}}(x)$, as
\begin{equation}
-b^{\sigma }\partial _{\sigma }\mathrm{{\mathcal{L}}}-\varepsilon ^{\sigma
\nu }x_{\nu }\partial _{\sigma }\mathrm{{\mathcal{L}}}=\frac{\partial
\mathrm{{\mathcal{L}}}}{\partial \Psi }\delta \Psi +\sum_{n\geq 1}\frac{%
\partial \mathrm{{\mathcal{L}}}}{\partial (\partial _{\mu _{1}}\ldots
\partial _{\mu _{n}}\Psi )}\partial _{\mu _{1}}\ldots \partial _{\mu
_{n}}\delta \Psi .  \label{Eq1119}
\end{equation}%
To derive the expression for the conserved current, one must use the
generalized higher-order Euler-Lagrange equation
\begin{equation}
\frac{\partial \mathrm{{\mathcal{L}}}}{\partial \Psi }+\sum_{n\geq
1}(-)^{n}\partial _{\mu _{n}}\ldots \partial _{\mu _{1}}\frac{\partial
\mathrm{{\mathcal{L}}}}{\partial (\partial _{\mu _{1}}\ldots \partial _{\mu
_{n}}\Psi )}=0.  \label{Eq1118}
\end{equation}%
By replacing the first term on the right-hand side of Eq.~(\ref{Eq1119}) with the
corresponding expression from the Euler-Lagrange equation (\ref{Eq1118}), one can rewrite
the right-hand side of Eq.~(\ref{Eq1119}) as follows:
\begin{equation}
\mathrm{r.h.s.}=-\sum_{n\geq 1}(-)^{n}\partial _{\mu _{n}}\ldots \partial
_{\mu _{1}}\frac{\partial \mathrm{{\mathcal{L}}}}{\partial (\partial _{\mu
_{1}}\ldots \partial _{\mu _{n}}\Psi )}\delta \Psi +\sum_{n\geq 1}\frac{%
\partial \mathrm{{\mathcal{L}}}}{\partial (\partial _{\mu _{1}}\ldots
\partial _{\mu _{n}}\Psi )}\partial _{\mu _{1}}\ldots \partial _{\mu
_{n}}\delta \Psi .  \label{Eq1120}
\end{equation}%
The purpose is to present the above expression in the form of the divergence
of some quantity. The $n$-th order term under the first summation symbol in
Eq.~(\ref{Eq1120}) can be rewritten in the form
\begin{equation} \label{n-term}
\begin{split}
&-(-)^{n}\partial _{\mu _{n}} \ldots \partial _{\mu _{1}}\frac{\partial
\mathrm{{\mathcal{L}}}}{\partial (\partial _{\mu _{1}}\ldots \partial _{\mu
_{n}}\Psi )}\delta \Psi   \\
=&-(-)^{n}\partial _{\mu _{n}}\left( \partial _{\mu _{n-1}}\ldots \partial
_{\mu _{1}}\frac{\partial \mathrm{{\mathcal{L}}}}{\partial (\partial _{\mu
_{1}}\ldots \partial _{\mu _{n}}\Psi )}\delta \Psi \right) \\
&-(-)^{n+1}\partial _{\mu _{n-1}}\ldots \partial _{\mu _{1}}\frac{\partial
\mathrm{{\mathcal{L}}}}{\partial (\partial _{\mu _{1}}\ldots \partial _{\mu
_{n}}\Psi )}\partial _{\mu _{n}}\delta \Psi .
\end{split}
\end{equation}%
The first term has the form of a divergence, whereas in the
second term, the derivative $\partial _{\mu _{n}}$ is shifted to the right
and acts on $\delta \Psi $. By rewriting the second term of (\ref{n-term}) in
the same way,
\begin{equation} \label{n-1-term}
\begin{split}
&-(-)^{n+1}\partial _{\mu _{n-1}}\ldots \partial _{\mu _{1}}\frac{\partial {%
\mathcal{L}}}{\partial (\partial _{\mu _{1}}\ldots \partial _{\mu _{n}}\Psi )%
}\partial _{\mu _{n}}\delta \Psi \\
=& -(-)^{n+1}\partial _{\mu _{n-1}}\left( \partial _{\mu _{n-2}}\ldots
\partial _{\mu _{1}}\frac{\partial {\mathcal{L}}}{\partial (\partial _{\mu
_{1}}\ldots \partial _{\mu _{n}}\Psi )}\partial _{\mu _{n}}\delta \Psi
\right) \\
&-(-)^{n+2}\partial _{\mu _{n-2}}\ldots \partial _{\mu _{1}}\frac{%
\partial {\mathcal{L}}}{\partial (\partial _{\mu _{1}}\ldots \partial _{\mu
_{n}}\Psi )}\partial _{\mu _{n-1}}\partial _{\mu _{n}}\delta \Psi,
\end{split}
\end{equation}%
we again obtain a divergence and one more derivative of $\delta\Psi$ in
the second term. This implies that through such recursion, one can shift the
derivative to the right until it lies immediately before $\delta \Psi $. With each such procedure,
the second term in the rewritten part of the expression changes sign and the first term has the form of a
divergence.

Finally, the last term in the recursion can be obtained by shifting
over $n$ derivatives; this term will have an additional sign $(-1)^{n}$, and
consequently, it will have a sign opposite to that of the second term of Eq.~(%
\ref{Eq1120}) and will therefore vanish.

Thus, the result of this procedure for the right-hand side of Eq.~(\ref{Eq1119}) has the form
\begin{equation}
\begin{split}
\mathrm{r.h.s.} &=  \sum_{k=1}^{n}(-)^{k+1}\partial _{\mu _{k}}\left( \partial
_{\mu _{k-1}}\ldots \partial _{\mu _{1}}\frac{\partial \mathrm{{\mathcal{L}}}%
}{\partial (\partial _{\mu _{1}}\ldots \partial _{\mu _{n}}\Psi )}\partial
_{\mu _{k+1}}\ldots \partial _{\mu _{n}}\delta \Psi \right) \\
&= \partial_{\sigma } \sum_{k=1}^{n}(-)^{k+1}\left( \partial _{\mu
_{k-1}}\ldots \partial _{\mu _{1}}\frac{\partial \mathrm{{\mathcal{L}}}}{%
\partial (\partial _{\mu _{1}}\ldots \partial _{\mu _{k-1}}\partial _{\sigma
}\partial _{\mu _{k+1}}\ldots \partial _{\mu _{n}}\Psi )}\partial _{\mu
_{k+1}}\ldots \partial _{\mu _{n}}\delta \Psi \right) .
\end{split}
\end{equation}

Finally, Eq.~(\ref{Eq1119}) can be written fully in the form of a divergence
as follows:
\begin{eqnarray}
\partial _{\sigma }[ \sum_{n\geq 1}\sum_{k=1}^{n}(-)^{k+1}\left(
\partial _{\mu _{k-1}} \,...\, \partial _{\mu _{1}}\frac{\partial \mathrm{{%
\mathcal{L}}}}{\partial (\partial _{\mu _{1}} \,...\, \partial _{\mu
_{k-1}}\partial _{\sigma }\partial _{\mu _{k+1}} \,...\, \partial_{\mu
_{n}}\Psi )}\partial_{\mu _{k+1}} \,...\, \partial _{\mu _{n}}\delta \Psi
\right)  \nonumber \\
\left. +b^{\sigma }\mathrm{{\mathcal{L}}}+\varepsilon^{\sigma \nu } x_{\nu }\mathrm{{\mathcal{L}}}\right] =0,  \label{Eq1121J}
\end{eqnarray}
where $\delta \Psi $ in the parentheses is given by Eq.~(\ref{delta-Psi}).
The terms that are linear in the parameter $\omega ^{a}$ determine the set of
conserved currents $\mathfrak{J}^{a\sigma }$ related to the internal
symmetry group. The terms that are proportional to the vector $-b^{\sigma }$
determine the conserved second-rank tensor that can be identified with the
energy-momentum tensor $\mathfrak{T}_{\mu }^{\sigma }$. Finally, the
terms that are proportional to the tensor $\varepsilon ^{\mu \nu }$ determine the
conserved third-rank tensor $\mathfrak{M}_{\mu \nu }^{\sigma }$. The spatial
components of this tensor correspond to the total angular
momentum density of the system. In the case of $n=1$, we obtain the standard results.
The set of conserved currents takes the form
\begin{equation} \label{Cons-J}
\begin{split}
\mathfrak{J}^{a\sigma } = \sum_{n\geq 1}\sum_{k=1}^{n}(-)^{k+1}\left(
\partial _{\mu _{k-1}}\ldots \partial _{\mu _{1}}\frac{\partial \mathrm{{%
\mathcal{L}}}}{\partial (\partial _{\mu _{1}}\ldots \partial _{\mu
_{k-1}}\partial _{\sigma }\partial _{\mu _{k+1}}\ldots \partial _{\mu
_{n}}\Psi )}\right) \\
\times \, \partial _{\mu _{k+1}}\ldots \partial _{\mu_{n}}(-iT^{a})\Psi (x),
\end{split}
\end{equation}
\begin{equation} \label{Cons-T}
\begin{split}
\mathfrak{T}_{\mu }^{\sigma } =\sum_{n\geq 1}\sum_{k=1}^{n}(-)^{k+1}\left(
\partial _{\mu _{k-1}}\ldots \partial _{\mu _{1}}\frac{\partial \mathrm{{%
\mathcal{L}}}}{\partial (\partial _{\mu _{1}}\ldots \partial _{\mu
_{k-1}}\partial _{\sigma }\partial _{\mu _{k+1}}\ldots \partial _{\mu
_{n}}\Psi )}\right) \\
\times \, \partial _{\mu _{k+1}}\ldots \partial
_{\mu _{n}} \partial_{\mu } \Psi (x)-\delta _{\mu }^{\sigma }\mathrm{{\mathcal{L}}},
\end{split}
\end{equation}
\begin{equation} \label{Cons-M}
\begin{split}
\mathfrak{M}_{\mu \nu }^{\sigma } =\sum_{n\geq
1}\sum_{k=1}^{n}(-)^{k+1}\left( \partial _{\mu _{k-1}}\ldots \partial _{\mu
_{1}}\frac{\partial \mathrm{{\mathcal{L}}}}{\partial (\partial _{\mu
_{1}}\ldots \partial _{\mu _{k-1}}\partial _{\sigma }\partial _{\mu
_{k+1}}\ldots \partial _{\mu _{n}}\Psi )}\right) \\
\times \partial _{\mu_{k+1}}\ldots \partial _{\mu _{n}}   (-i) \left( \Lambda_{\mu \nu} + \Sigma
_{\mu \nu }\right) \Psi (x)-\left( x_{\mu }\delta _{\nu }^{\sigma }-x_{\nu
}\delta _{\mu }^{\sigma }\right) \mathrm{{\mathcal{L}}}.
\end{split}
\end{equation}
Noether's theorem allows us to find the conserved currents accurately to
within an arbitrary factor. In Eqs.~(\ref{Cons-J}) - (\ref{Cons-M}), the
factors are chosen in such a way that the quantity $\mathfrak{T}_{0}^{0}$
coincides with the energy density defined by the
Legendre transform of the Lagrangian. The quantity $\mathfrak{M}%
_{\alpha \beta }^{0}$ then coincides with the angular
momentum density of the system for the spatial indices $\alpha $ and $\beta $.
Equation (\ref{Cons-T}) is in agreement with Ref.~\cite{Szabados2009}.

In a non-local field theory, we expand non-local operators of the Lagrangian in an infinite power series
over the differential operators. The conserved currents are then given by Eqs.~(\ref{Cons-J}) - (\ref{Cons-M}),
with the summation over $n$ extended to $+\infty$.
This method is applied below to construct the conserved currents in a non-local charged scalar field theory.

\section{Non-local charged scalar field}
\renewcommand{\theequation}{3.\arabic{equation}}
\setcounter{equation}{0}
$\;$ \vspace{-12pt}

We consider an example of a non-local charged scalar field described by the Lagrangian
\begin{equation}
\mathrm{{\mathcal{L}}}=\frac{1}{2}\phi ^{\ast }\left( i\partial _{t}-\sqrt{-\Delta +m^{2}}\right) \phi +\mathrm{c.c.}  \label{par-Lagr}
\end{equation}
The particles follow a relativistic dispersion law $E(\mathbf{p})=\sqrt{%
\mathbf{p}^{2}+m^{2}}$. Because of the absence of negative-frequency
solutions, the particles do not have antiparticles.

Let us check whether CPT invariance holds in the non-local field theory defined by (\ref{par-Lagr}).
First consider the charge-conjugation operation, C.
In the momentum space given by $p^\alpha = - i \nabla = -(p^{\alpha})^{*}$, with $\alpha = 1, 2, 3$, we replace the particles' momenta
in (\ref{par-Lagr}) with the generalized momenta, $p_\mu \rightarrow p_\mu - e A_\mu$. The evolution equation in an external
electromagnetic field takes the form
\beq (i \partial_t - e A_0) \phi = \sqrt{ (\mathbf{p} - e \mathbf{A})^2 + m^2 } ~\phi. \label{cC} \eeq
For the complex conjugate scalar field, we obtain
\beq (i \partial_t + e A_0) \phi^* = - \sqrt{ (\mathbf{p} - e \mathbf{A})^2 + m^2 } ~\phi^*. \label{ccC} \eeq
Together with the sign reversal of the charge $e$ in Eq.~(\ref{ccC}), a negative sign appears at the root.
Obviously, the charge-conjugation symmetry is broken. Violation of the C symmetry means that the properties
of a particle and its corresponding antiparticle are different or, as in our case, the corresponding antiparticles do not exist.

One can easily check that the Lagrangian given in (\ref{par-Lagr}) is invariant under the parity transformation P:
$\phi (t, \textbf{x}) \to \phi (t, - \textbf{x})$. By the same analysis, one can check that the time-reversal symmetry, T:
$\phi(t, \textbf{x}) \rightarrow \phi^{*}( - t, \textbf{x})$,  is conserved as well. Thus, the Lagrangian of (\ref{par-Lagr})
is symmetric under P and T transformations, whereas the C symmetry is broken. The combined CPT symmetry is therefore broken,
which is consistent with the fact that the theory is non-local.

The Lagrangian expressed in (\ref{par-Lagr}) is explicitly invariant under global phase rotations of $\phi $, which may imply
the existence of a conserved vector current. The Lagrangian given in (\ref{par-Lagr}) is also explicitly invariant under
space-time translations and three-dimensional rotations. We thus expect the existence of conserved energy-momentum and angular momentum tensors.
The dispersion law takes a relativistic form; therefore, the field theory of (\ref{par-Lagr}) is apparently invariant under
boost transformations. This symmetry is, however, implicit, and we do not discuss its consequences here.
We thus restrict ourselves to the case of $\epsilon^{0\alpha} = 0$, $\epsilon^{\alpha \beta} \neq 0$.

We will work in terms of a power series over the derivatives. Expanding $\mathcal{L}$, one can rewrite it as follows:
\begin{equation}
\mathrm{{\mathcal{L}}}=\frac{1}{2}\left( \phi ^{\ast }i\partial _{t}\phi
-\sum_{l=0}^{\infty }f_{l}(m)\phi ^{\ast }\Delta ^{l}\phi \right) +\mathrm{%
c.c.},  \label{Lagr-series}
\end{equation}%
where
\begin{equation}
f_{l}(m)=(-1)^{l}\frac{\Gamma (\frac{3}{2})}{l!~\Gamma (\frac{3}{2}%
-l)~m^{2l-1}},  \label{flm}
\end{equation}%
such that%
\begin{equation}
\sum_{l=0}^{\infty }f_{l}(m)x ^{l}=\sqrt{m^{2}-x }. \label{summ}
\end{equation}

\subsection{Time-like components} \label{tcomponents}
$\;$ \vspace{-12pt}

One can easily find the zeroth component of the conserved currents because the
Lagrangian expressed in (\ref{Lagr-series}) contains only the first derivative with
respect to time and there are no mixed derivatives. This implies that the
series in Eqs.~(\ref{Cons-J}) - (\ref{Cons-M}) are truncated at the first term
of the sum. Thus, the charge density $\mathfrak{J}^{0}$, the energy density $%
\mathfrak{T}_{\mu }^{0}$ and the angular momentum density $\mathfrak{M}_{\alpha \beta }^{0}$
take the following simple forms:
\begin{eqnarray}
\mathfrak{J}^{0} &=&\phi ^{\ast }\phi ,  \label{JJ0} \\
\mathfrak{T}_{\mu }^{0} &=&\frac{1}{2}\phi ^{\ast }i\overleftrightarrow{%
\partial }_{\mu }\phi ,  \label{TT0} \\
\mathfrak{M}_{\alpha \beta }^{0} &=&\frac{1}{2}\phi ^{\ast } \mathcal{R}_{\alpha \beta} \phi
+ \frac{1}{2} (\mathcal{R}_{\alpha \beta} \phi )^{\ast }  \phi,  \label{MM0}
\end{eqnarray}%
where $\overleftrightarrow{\partial }_{\mu }=\overrightarrow{\partial }_{\mu
}-\overleftarrow{\partial }_{\mu }$ and $\mathcal{R}_{\alpha \beta}$ is defined following Eq.~(\ref{delta-Psi}).

We turn to momentum space,
substituting into Eqs.~(\ref{JJ0})-(\ref{MM0}) plane waves for outgoing and incoming particles with momenta $p^{\prime }$ and $p$. The
four-momentum operator in coordinate space is given by
$
p^{\mu }=(E,\mathbf{p})=\left( i\partial _{t},-i\nabla \right) .
$
In terms of the transition matrix elements, the conserved currents (\ref{JJ0})-(\ref{MM0}) take the forms
\begin{eqnarray}
\mathfrak{J}^{0}(p^{\prime },p) &=&1,  \label{555} \\
\mathfrak{T}_{\mu }^{0}(p^{\prime },p) &=&\frac{1}{2}(p^{\prime }+p)_{\mu },
\label{55} \\
\mathfrak{M}_{\alpha \beta }^{0}(p^{\prime },p) &=&\frac{1}{2}\left( \hat{x}_{\alpha
}(p^{\prime }+p)_{\beta }-\hat{x}_{\beta }(p^{\prime }+p)_{\alpha  }\right) ,
\label{5}
\end{eqnarray}%
where $\hat{x}_{\mu }=-i\partial /\partial p^{\mu }=i\partial /\partial
p^{\prime \mu }$. 

To find the spatial components of the conserved currents, one must specify the action
of the derivatives in expressions (\ref{Cons-J}) - (\ref{Cons-M}).
The rules that are useful for deriving the expressions for the conserved currents are given in Appendix~\ref{appen1}.

\subsection{Vector current} \label{sec-current}
$\;$ \vspace{-12pt}

Following the rules listed in Appendix~\ref{appen1}, we find the spatial components of
the vector current as follows:
\begin{equation} \label{Eq1124}
\begin{split}
{\mathfrak{J}}^{\alpha } =&\frac{i}{2}\sum_{l=1}^{\infty
}f_{l}(m)\sum_{k=1,3,5,...}^{2l-1}(\partial _{\alpha _{k-1}}\ldots \partial
_{\alpha _{1}}\phi ^{\ast })\delta ^{\alpha _{1}\alpha _{2}}\ldots
\delta ^{\alpha \alpha _{k+1}}\ldots \delta
^{\alpha _{2l-1}\alpha _{2l}} \partial _{\alpha _{k+1}}\ldots \partial_{\alpha _{2l}}\phi   \\
-&\frac{i}{2}\sum_{l=1}^{\infty }f_{l}(m)\sum_{k=2,4,6,...}^{2l}(\partial
_{\alpha _{k-1}}\ldots \partial _{\alpha _{1}}\phi ^{\ast })\delta ^{\alpha
_{1}\alpha _{2}}\ldots \delta^{\alpha _{k-1}\alpha }
\ldots \delta ^{\alpha _{2l-1}\alpha _{2l}}\partial
_{\alpha _{k+1}}\ldots \partial _{\alpha _{2l}}\phi \\
&+\mathrm{c.c.,}
\end{split}
\end{equation}
where $f_{l}(m)$ is given by Eq.~(\ref{flm}) and $\delta \phi =-i\phi $ and $%
\delta \phi ^{\ast }=i\phi ^{\ast }$ for the $U(1)$ symmetry group.

Let us write Eq.~(\ref{Eq1124}) in the lowest-order approximation. Equation (\ref{flm}) yields $f_{1}(m)=-1/(2m)$. The space-like component of the vector current ${\mathfrak{J}}^{\sigma }$ reduces to the standard expression
\begin{equation} \label{J1order}
{\mathfrak{J}}^{\alpha } = \frac{1}{2m}\phi ^{\ast }i\overset{%
\leftrightarrow }{\partial ^{\alpha }}\phi + \ldots.
\end{equation}

By performing contractions of the indices and with the aid of Eq.~(\ref{Eqap5}) from Appendix~\ref{appen1}, we obtain
\begin{equation} \label{1111}
\begin{split}
{\mathfrak{J}}^{\alpha } = &-\frac{i}{2}\sum_{l=1}^{\infty
}f_{l}(m)\sum_{k=1,3,5,...}^{2l-1}(\bigtriangleup ^{(k-1)/2}\phi ^{\ast
})\bigtriangleup ^{l-(k+1)/2}\partial ^{\alpha }\phi   \\
&+\frac{i}{2}\sum_{l=1}^{\infty
}f_{l}(m)\sum_{k=2,4,6,...}^{2l}(\bigtriangleup ^{(k-2)/2}\partial ^{\alpha
}\phi ^{\ast })\bigtriangleup ^{l-k/2}\phi +\mathrm{c.c.}
\end{split}
\end{equation}
The sum of the first two terms is real, so adding the complex conjugate expression doubles the result.
After some simple algebra and with the use of Eq.~(\ref{factor-ser}), we obtain
\begin{equation} \label{3333}
{\mathfrak{J}}^{\alpha } =  \phi ^{\ast }i{\mathcal{D}}^{\alpha }
\phi ,
\end{equation}
where
\begin{equation}
{\mathcal{D}}^{\alpha } = \frac{\overleftrightarrow{\partial }^{\alpha }}{\sqrt{m^{2}-(\overleftarrow{\bigtriangleup })}+\sqrt{m^{2}-(\overrightarrow{\bigtriangleup })}}.
\label{cal-D}
\end{equation}
The detailed derivation of Eq.~(\ref{3333}) is given in Appendix~\ref{appen2}.
In terms of the four-dimensional operator $i\mathcal{D}^{\sigma } \equiv (1,i\mathcal{D}^{\alpha})$, the four-dimensional vector current can be written as
\begin{equation} \label{cov-D}
{\mathfrak{J}}^{\sigma } =  \phi ^{\ast }i{\mathcal{D}}^{\sigma } \phi.
\end{equation}

It is useful to rewrite the vector current in the momentum space. By substituting the plane waves $\phi^{*}(x) \sim e^{ip^{\prime} x}$ and $\phi(x) \sim e^{-ip x}$ with momenta $p^{\prime}$ and $p$ into Eq.~(\ref{cov-D}) and omitting the exponential factors from the final expression, we obtain
\begin{equation}
{\mathfrak{J}}^{\sigma }(p^{\prime },p)=\left( 1,\frac{\mathbf{p}^{\prime }+\mathbf{p}}{E(\mathbf{p}^{\prime })+E(\mathbf{p})}\right) .
\label{Eq1131bis}
\end{equation}%
On the mass shell, the vector current is conserved:
\begin{equation}
\partial_{\sigma}{\mathfrak{J}}^{\sigma } =0. \label{JJp-Cons}
\end{equation}%

The variational derivative of the action functional $S=\int d^{4}x\mathrm{{\mathcal{L}}}$ with respect to the vector field $A_{\sigma}(x) $,
\begin{equation}
\mathfrak{J}^{\sigma }(x)=-\frac{\delta S}{\delta A_{\sigma}(x)},
\label{off-ms}
\end{equation}
introduced into $\mathrm{\mathcal{L}}$ with the use of minimal substitution,
is associated for $A_{\sigma}(x) = 0$ with a vector current.
Current (\ref{off-ms}) is defined off the mass shell and
it coincides with the Noether current (\ref{cov-D}) on the mass shell, as shown in Appendix~\ref{appenb}.
Multiplying (\ref{Eq1131bis}) with $(p^{\prime }-p)_{\sigma }$ yields the result
\begin{equation}
(p^{\prime }-p)_{\sigma }{\mathfrak{J}}^{\sigma }(p^{\prime },p) = G^{-1}(p^{\prime})-G^{-1}(p),
\label{Eq1131bisbis}
\end{equation}%
where $G(p)=(p^{0}-\sqrt{m^{2}+\mathbf{p}^{2}})^{-1}$ is the particle propagator.
This equation can be recognized as the Ward identity.

The field $\phi_s(x)$, which behaves like a true scalar under Lorentz transformations, may be defined by the equation
$\phi(x) = \left( i\partial _{t} + \sqrt{-\Delta + m^{2}}\right)^{1/2} \phi_{s}(x)$.
In terms of $\phi_{s}(x)$, the Lagrangian (\ref{par-Lagr}) takes the explicitly covariant form
$\mathrm{{\mathcal{L}}}=\frac{1}{2}\phi_s ^{\ast }( - \square - m^2) \phi_s +\mathrm{c.c.}$
The non-local operator $\left( i\partial _{t} + \sqrt{-\Delta + m^{2}}\right)^{1/2}$ eliminates from $\phi_s(x)$ the negative-frequency solutions.
Since the proper Lorentz transformations do not mix plane waves with the positive and negative frequencies,
the classical non-local field theory (\ref{par-Lagr}) appears to be Lorentz covariant.
The interaction preserving the covariance can be introduced, e.g., by adding to $\mathrm{{\mathcal{L}}}$ the term $\lambda |\phi_s|^4$.

Equations~(\ref{Cons-J}) - (\ref{Cons-M}) are straightforward generalizations of the Noether currents of a local field theory.
Noether's theorem applied to $\mathrm{{\mathcal{L}}}(\phi_{s}(x))$ leads, however, to conserved currents
that differ from those of Eqs.~(\ref{Cons-J}) - (\ref{Cons-M}).
A family of the conserved currents apparently exists when non-local field transformations are permitted.
The conserved vector current of $\mathrm{{\mathcal{L}}}(\phi_{s}(x))$, e.g., takes the form
${(p^{\prime }+ p)^{\sigma}}/
{\sqrt{{{2E(\mathbf{p}^{\prime })2E(\mathbf{p})}}} }.
$
Among the conserved currents, the expression (\ref{cov-D}) is highlighted by the coincidence with (\ref{off-ms}).

\subsection{Energy-momentum tensor}
$\;$ \vspace{-12pt}

An analysis that is fundamentally identical to that presented in the previous section leads to the conserved energy-momentum tensor.
Considering that $\delta _{\mu }^{\alpha }\mathcal{L}=0$ for the fields that satisfy the
equations of motion, one can rewrite Eq.~(\ref{Cons-T}) with the Lagrangian given in (\ref{par-Lagr}) in the form
\begin{equation}
\begin{split}
{\mathfrak{T}}_{\mu }^{\alpha } =&-\frac{1}{2}\sum_{l=1}^{\infty
}f_{l}(m)\sum_{k=1,3,5,...}^{2l-1} (\partial _{\alpha _{k-1}}\ldots
\partial _{\alpha _{1}}\phi ^{\ast })\delta ^{\alpha _{1}\alpha _{2}}\ldots
\delta ^{\alpha \alpha _{k+1}}\ldots
\delta ^{\alpha _{2l-1}\alpha _{2l}} \partial _{\alpha
_{k+1}}\ldots \partial _{\alpha _{2l}} \partial_{\mu } \phi   \\
&+\frac{1}{2}\sum_{l=1}^{\infty
}f_{l}(m)\sum_{k=2,4,6,...}^{2l}(\partial _{\alpha _{k-1}}\ldots
\partial _{\alpha _{1}}\phi ^{\ast })\delta ^{\alpha _{1}\alpha _{2}}\ldots
\delta ^{\alpha _{k-1}\alpha }
\ldots
\delta ^{\alpha _{2l-1}\alpha _{2l}} \partial _{\alpha
_{k+1}}\ldots \partial _{\alpha _{2l}} \partial_{\mu } \phi  \\
&+ \mathrm{c.c.}. \label{Cons-T-delta}
\end{split}
\end{equation}
The lowest-order $l=1$ term of the expansion yields
\begin{equation}
{\mathfrak{T}}_{\mu }^{\alpha } = \frac{1}{4m}\phi^{\ast }i\overset{\leftrightarrow }{\partial^{\alpha }}
i\overset{\leftrightarrow }{\partial _{\mu }}\phi + \ldots.
\end{equation}
By performing contractions of the indices in Eq.~(\ref{Cons-T-delta}), we obtain
\begin{equation} \label{1111T}
\begin{split}
{\mathfrak{T}}_{\mu}^{\alpha } =& - \frac{1}{2}\sum_{l=1}^{\infty
}f_{l}(m)\sum_{k=1,3,5,...}^{2l-1}(\bigtriangleup ^{(k-1)/2}\phi ^{\ast
})\bigtriangleup ^{l-(k+1)/2}\partial^{\alpha }\partial_{\mu}\phi   \\
&+\frac{1}{2}\sum_{l=1}^{\infty
}f_{l}(m)\sum_{k=2,4,6,...}^{2l}(\bigtriangleup ^{(k-2)/2}\partial^{\alpha
}\phi ^{\ast })\bigtriangleup ^{l-k/2}\partial_{\mu}\phi +\mathrm{c.c.}
\end{split}
\end{equation}

Using Eq.~(\ref{factor-ser}), the summation in Eq.~(\ref{1111T}) can be performed in the same way as for the conserved current.
The energy-momentum tensor finally takes the form
\begin{equation} \label{enmomten}
\mathfrak{T}_{\mu }^{\sigma }= \frac{1}{2}\phi ^{\ast }
i\mathcal{D}^{\sigma} i\overleftrightarrow{\partial }_{\mu }\phi .
\end{equation}
A detailed derivation of this expression for $\sigma = 1,2,3$ is given in Appendix~\ref{appen2}.
Equation~(\ref{enmomten}) defines four conserved quantities, one for each component of the translation parameter $b^{\sigma }$.
In momentum space,
\begin{equation} \label{offT}
\mathfrak{T}_{\mu }^{\sigma }(p^{\prime },p)
=\mathfrak{J}^{\sigma }(p^{\prime },p)\frac{1}{2}(p^{\prime }+p)_{\mu },
\end{equation}
where $\mathfrak{J}^{\sigma }(p^{\prime },p)$ is given by Eq.~(\ref{Eq1131bis}). Using Eq.~(\ref{Eq1131bisbis}),
we obtain the conservation condition for the energy-momentum tensor on the mass shell:
\begin{equation}
\partial_{\sigma}{\mathfrak{T}}_{\mu }^{\sigma } =0. \label{TTp-Cons}
\end{equation}%

\subsection{Angular momentum tensor}
%
$\;$ \vspace{-12pt}

The conservation of angular momentum arises from the invariance of the system with respect to rotation. Taking $\Sigma _{\mu \nu }=0$
for the charged scalar field and substituting $\delta _{\mu }^{\alpha }\mathcal{L}=0$ into Eq.~(\ref{Cons-M}), one can write the expression for
the angular momentum density in the following form:
\begin{eqnarray} \label{Cons-M-delta}
{\mathfrak{M}}_{\alpha \beta }^{\gamma } &=& \frac{i}{2}\sum_{l=1}^{\infty
}f_{l}(m)\sum_{k=1,3,5,...}^{2l-1}(\partial _{\alpha _{k-1}}\ldots \partial
_{\alpha _{1}}\phi ^{\ast })\delta ^{\alpha _{1}\alpha _{2}}\ldots
\delta ^{\gamma \alpha _{k+1}}\ldots \delta
^{\alpha _{2l-1}\alpha _{2l}} \partial _{\alpha _{k+1}}\ldots \partial_{\alpha _{2l}}\mathcal{R}_{\alpha \beta}\phi \nonumber \\
&-&\frac{i}{2}\sum_{l=1}^{\infty }f_{l}(m)\sum_{k=2,4,6,...}^{2l}(\partial
_{\alpha _{k-1}}\ldots \partial _{\alpha _{1}}\phi ^{\ast })\delta ^{\alpha
_{1}\alpha _{2}}\ldots \delta^{\alpha _{k-1}\gamma }
\ldots \delta ^{\alpha _{2l-1}\alpha _{2l}}\partial
_{\alpha _{k+1}}\ldots \partial _{\alpha _{2l}} \mathcal{R}_{\alpha \beta}\phi \nonumber \\
&+&\mathrm{c.c.}
\end{eqnarray}
The first terms of the series expansion are
\begin{equation}
{\mathfrak{M}}_{\alpha \beta }^{\gamma } =\frac{1}{4m}
 \phi^{\ast }i\overset{\leftrightarrow }{\partial
^{\gamma }}  \mathcal{R}_{\alpha \beta}\phi + \frac{1}{4m}
(\mathcal{R}_{\alpha \beta}\phi)^{\ast} i\overset{\leftrightarrow }{\partial^{\gamma }} \phi
+ \ldots .
\end{equation}
By performing contraction of the indices in Eq.~(\ref{Cons-M-delta}),
we obtain
\begin{equation} \label{1111M}
\begin{split}
\MM_{\alpha \beta }^{\gamma } = &  \frac{i}{2}\sum_{l=1}^{\infty
}f_{l}(m)\sum_{k=1,3,5,...}^{2l-1}(\bigtriangleup ^{(k-1)/2}\phi ^{\ast
})\bigtriangleup ^{l-(k+1)/2}\partial^{\gamma } \mathcal{R}_{\alpha \beta}\phi   \\
-&\frac{i}{2}\sum_{l=1}^{\infty
}f_{l}(m)\sum_{k=2,4,6,...}^{2l}(\bigtriangleup ^{(k-2)/2}\partial^{\gamma
}\phi ^{\ast })\bigtriangleup ^{l-k/2} \mathcal{R}_{\alpha \beta}\phi +\mathrm{c.c.}
\end{split}
\end{equation}
The arguments presented in Appendix~\ref{appen2} enable the summation of the series in Eq.~(\ref{1111M}), yielding
\begin{equation} \label{MPD11}
\mathfrak{M}_{\alpha \beta }^{\sigma }= \frac{1}{2}\phi ^{\ast } i\mathcal{D}^{\sigma}
\mathcal{R}_{\alpha \beta}\phi + \frac{1}{2}(\mathcal{R}_{\alpha \beta}\phi)^{\ast }i\mathcal{D}^{\sigma}
\phi.
\end{equation}
For $\sigma = 0$, we recover Eq.~(\ref{MM0}). $\mathcal{R}_{\alpha \beta}$ is not diagonal in the momentum representation, so the momentum-space representation of $\mathfrak{M}_{\alpha \beta }^{\gamma }$ offers no significant advantages. Using the equations of motion, one can verify that
\begin{equation} \label{MMp-Cons}
\partial_{\sigma} \mathfrak{M}_{\alpha \beta }^{\sigma } = 0.
\end{equation}

The conserved currents defined by Eq.~(\ref{MPD11}) correspond to the space-like components of the parameter $\varepsilon^{\alpha \beta}$,
which describe a rotation; thus, the conserved charges are the components of the angular momentum tensor.

\section{Conclusion}
\renewcommand{\theequation}{4.\arabic{equation}}
\renewcommand{\thesection}{}
\setcounter{equation}{0}
$\;$ \vspace{-12pt}

In non-local field theory with an internal symmetry and symmetries of the Poincar\'e group
there exist conserved vector current and energy-momentum and angular momentum tensors. Expressions (\ref{Cons-J}) - (\ref{Cons-M})
solve explicitly the problem of finding the corresponding Noether currents in terms of infinite series of the field's derivatives.

Equaitons (\ref{Cons-J}) - (\ref{Cons-M}) were used for the construction of the conserved currents in a non-local theory of a charged scalar field
with explicit symmetries of phase rotations, translations and spatial rotations. 
Using combinatorial arguments, we summed the infinite series over derivatives of fields
and obtained the simple analytical expressions (\ref{cov-D}), (\ref{enmomten}) and (\ref{MPD11}) for the corresponding Noether currents.

\vspace{15pt}
One of the authors (A.A.T.) wishes to acknowledge the kind hospitality of Bogoliubov Laboratory of Theoretical Physics.
This work was supported in part by RFBR Grant No.~16-02-01104, Grant~No.~HLP-2015-18~of the Heisenberg-Landau~Program and MEYS Grant No. LG14004.

\begin{appendices}

\section{Field derivatives} \label{appen1}
\renewcommand{\theequation}{A.\arabic{equation}}
\setcounter{equation}{0}
$\;$ \vspace{-12pt}

In this section, we consider algebraic rules for the manipulation of the field's derivatives in Minkowski space.
The proofs are valid, however, in the general case of $\mathbb{R}^{m,n}$.
The fields $ \Psi $ and their derivatives are not assumed to be smooth; therefore, the sequence of the differentiation operations matters.
As a result,
\begin{equation*}
\frac{\partial \mathrm{{\mathcal{L}}}}{\partial (\partial _{\mu
_{1}}\partial _{\mu _{2}}\ldots \partial _{\mu _{n}}\Psi )}
\end{equation*}%
is not necessarily symmetric under the permutation of indices.
The conserved currents (\ref{Cons-J})-(\ref{Cons-M}) are then calculated using the
following formulas:
\begin{eqnarray}
{\frac{\partial }{\partial (\partial _{\mu }\Psi )}\partial _{\tau }\Psi }
&=&{\delta _{\tau }^{\mu }}, \\
{\frac{\partial }{\partial (\partial _{\mu }\partial _{\nu }\Psi )}\partial
_{\tau }\partial _{\sigma }\Psi } &=&{\delta _{\tau }^{\mu }\delta _{\sigma
}^{\nu }}, \\
& \vdots & \nonumber \\
{\frac{\partial }{\partial (\partial _{\mu _{1}}\partial _{\mu _{2}}\ldots
\partial _{\mu _{n}}\Psi )}\partial _{\nu _{1}}\partial _{\nu _{2}}\ldots
\partial _{\nu _{n}}\Psi } &=&{\delta _{\nu _{1}}^{\mu _{1}}\delta _{\nu
_{2}}^{\mu _{2}}\ldots \delta _{\nu _{n}}^{\mu _{n}}}.
\end{eqnarray}%
In particular,
\begin{equation}
\frac{\partial }{\partial (\partial _{\mu }\partial _{\nu }\Psi )}{\square }%
\Psi =g^{\mu \nu }  \label{gmunu}
\end{equation}%
and
\begin{equation} \label{Eqap5}
(\partial _{\mu _{1}}\partial _{\mu _{2}}\ldots \partial _{\mu _{2l}}\Psi )%
\frac{\partial }{\partial (\partial _{\mu _{1}}\partial _{\mu _{2}}\ldots
\partial _{\mu _{2l}}\Psi )}{\square }^{l}\Psi ={\square }^{l}\Psi .
\end{equation}%

After the replacements ${\square }\rightarrow \Delta$ and $g^{\mu \nu }\rightarrow
\delta ^{\alpha \beta }$, these formulas also hold in Euclidean space.
The Euclidean versions of Eqs. (A.1) - (A.5) were used to derive Eqs.~(\ref{Eq1124}), (\ref{Cons-T-delta}) and (\ref{Cons-M-delta}).

\section{Series summation} \label{appen2}
\renewcommand{\theequation}{B.\arabic{equation}}
\setcounter{equation}{0}
$\;$ \vspace{-12pt}


The factor $1/2$ in Eq.~(\ref{1111}) vanishes with the addition of the complex conjugate part. The result can be written in the form
\begin{eqnarray}
{\mathfrak{J}}^{\alpha } &=&i\sum_{l=1}^{\infty
}f_{l}(m)\sum_{k=1,3,5,...}^{2l-1}(\bigtriangleup ^{(k-1)/2}\phi ^{\ast
})\bigtriangleup ^{l-(k+1)/2}\partial^{\alpha }\phi   \nonumber \\
&-&i\sum_{l=1}^{\infty }f_{l}(m)\sum_{k=2,4,6,...}^{2l}(\bigtriangleup
^{(k-2)/2}\partial ^{\alpha }\phi ^{\ast })\bigtriangleup ^{l-k/2}\phi  \nonumber \\
&=&i\sum_{l=1}^{\infty }f_{l}(m)\phi ^{\ast}
\left( (\overrightarrow{%
\bigtriangleup })^{l-1} + (\overleftarrow{%
\bigtriangleup })(\overrightarrow{\bigtriangleup })^{l-2} +...+ (\overleftarrow{\bigtriangleup })^{l-1} \right) \partial^{\alpha }\phi \nonumber  \\
&-&i\sum_{l=1}^{\infty }f_{l}(m) \partial^{\alpha }\phi ^{\ast } \left(
(\overrightarrow{\bigtriangleup })^{l-1} + (\overleftarrow{\bigtriangleup })(\overrightarrow{\bigtriangleup }%
)^{l-2} +...+ (\overleftarrow{%
\bigtriangleup })^{l-1} \right) \phi. \nonumber \\
&=&i\sum_{l=1}^{\infty }f_{l}(m) \phi ^{\ast } \left(
(\overrightarrow{\bigtriangleup })^{l-1} + (\overleftarrow{\bigtriangleup })(\overrightarrow{\bigtriangleup }%
)^{l-2} +...+ (\overleftarrow{%
\bigtriangleup })^{l-1} \right) \overleftrightarrow{\partial }^{\alpha }\phi. \label{B2}
\end{eqnarray}
The arrows over $\bigtriangleup$ indicate the direction in which the differentiation acts.
The series can be summed up using the factorization formula
\begin{equation}
x^{l}-y^{l}=(x-y)\sum_{k=1}^{l}x^{k-1}y^{l-k}.  \label{factor-ser}
\end{equation}%
This formula allows the simplification of the expression in brackets of Eq.~(\ref{B2}):
\begin{eqnarray}
{\mathfrak{J}}^{\alpha }
&=&i\sum_{l=1}^{\infty }f_{l}(m)\phi ^{\ast }\frac{1}{(\overrightarrow{%
\bigtriangleup })-(\overleftarrow{\bigtriangleup })}[(\overrightarrow{%
\bigtriangleup })^{l}-(\overleftarrow{\bigtriangleup })^{l}]%
\overleftrightarrow{\partial }^{\alpha }\phi  \nonumber \\
&=&i\sum_{l=0}^{\infty }f_{l}(m)\phi ^{\ast }\frac{1}{(\overrightarrow{%
\bigtriangleup })-(\overleftarrow{\bigtriangleup })}[(\overrightarrow{%
\bigtriangleup })^{l}-(\overleftarrow{\bigtriangleup })^{l}]%
\overleftrightarrow{\partial }^{\alpha }\phi  \nonumber \\
&=&i\phi ^{\ast }\frac{1}{(\overrightarrow{\bigtriangleup })-(%
\overleftarrow{\bigtriangleup })}[\sqrt{m^{2}-(\overrightarrow{%
\bigtriangleup })}-\sqrt{m^{2}-(\overleftarrow{\bigtriangleup })}]%
\overleftrightarrow{\partial }^{\alpha }\phi  \nonumber \\
&=&\phi ^{\ast }\frac{i\overleftrightarrow{\partial }^{\alpha }}{\sqrt{%
m^{2}-(\overrightarrow{\bigtriangleup })}+\sqrt{m^{2}-(\overleftarrow{%
\bigtriangleup })}}\phi . \label{eqtrans}
\end{eqnarray}
In the transition to the second line, we use the fact that $\bigtriangleup^0 = 1$.
The third line is obtained with the aid of Eq.~(\ref{summ}). We thus arrive at Eq.~(\ref{3333}).

On the way we proved a useful formula
\begin{equation}
\sum_{l=1}^{\infty}f_{l}(m)\sum_{k=1}^{l}x^{k-1}y^{l-k} = - \frac{1}{\sqrt{m^2-x} + \sqrt{m^2-y}}.
\label{binom}
\end{equation}

The sum over $k$ in Eq.~(\ref{1111T}) can be written explicitly as follows:
\begin{eqnarray}
{\mathfrak{T}}_{\mu}^{\alpha } &=&
\frac12 \sum_{l=1}^{\infty }f_{l}(m)\phi ^{\ast} \left( (\overrightarrow{%
\bigtriangleup })^{l-1} + (\overleftarrow{%
\bigtriangleup })(\overrightarrow{\bigtriangleup })^{l-2} +...+ (\overleftarrow{\bigtriangleup })^{l-1} \right) \partial^{\alpha } \partial_{\mu }\phi \nonumber  \\
&-& \frac12 \sum_{l=1}^{\infty }f_{l}(m) \partial^{\alpha }\phi ^{\ast } \left(
(\overrightarrow{\bigtriangleup })^{l-1} + (\overleftarrow{\bigtriangleup })(\overrightarrow{\bigtriangleup }%
)^{l-2} +...+ (\overleftarrow{%
\bigtriangleup })^{l-1} \right) \partial_{\mu} \phi + \mathrm{c.c.}
\end{eqnarray}
Here, the indices are those of tensors in Minkowski space (e.g., $\partial^{\alpha } = - \partial_{\alpha }$).
Using the formula given in (\ref{binom}),
the energy-momentum tensor can be found to be
\begin{equation}
{\mathfrak{T}}_{\mu}^{\alpha } = \frac12 \phi ^{\ast }\frac{ i\overleftrightarrow{\partial }^{\alpha } i\overleftrightarrow{\partial }_{\mu }}{\sqrt{%
m^{2}-(\overrightarrow{\bigtriangleup })}+\sqrt{m^{2}-(\overleftarrow{%
\bigtriangleup })}}\phi .
\end{equation}

Equation (\ref{1111M}) leads to
\begin{eqnarray}
{\mathfrak{M}}_{\alpha\beta}^{\gamma } &=& - \frac{i}{2} \sum_{l=1}^{\infty }f_{l}(m)\phi ^{\ast}
\left( (\overrightarrow{%
\bigtriangleup })^{l-1} + (\overleftarrow{%
\bigtriangleup })(\overrightarrow{\bigtriangleup })^{l-2} +...+ (\overleftarrow{\bigtriangleup })^{l-1} \right) \partial^{\gamma } \RR_{\alpha\beta} \phi  \nonumber  \\
&+& \frac{i}{2} \sum_{l=1}^{\infty }f_{l}(m) \partial^{\gamma }\phi ^{\ast } \left(
(\overrightarrow{\bigtriangleup })^{l-1} + (\overleftarrow{\bigtriangleup })(\overrightarrow{\bigtriangleup }%
)^{l-2} +...+ (\overleftarrow{%
\bigtriangleup })^{l-1} \right) \RR_{\alpha\beta} \phi  + \mathrm{c.c.}
\end{eqnarray}
By writing the complex conjugate part of the expression explicitly, one can simplify the above equation using the formula as expressed in (\ref{binom}):
\begin{eqnarray}
{\mathfrak{M}}_{\alpha\beta}^{\gamma} &=&\frac{1}{2} \phi^{\ast } \frac{ i \overleftrightarrow{\partial }^{\gamma } }{\sqrt{%
m^{2}-(\overrightarrow{\bigtriangleup })}+\sqrt{m^{2}-(\overleftarrow{%
\bigtriangleup })}} \RR_{\alpha\beta}\phi \nonumber \\
&+&
\frac{1}{2} (\RR_{\alpha\beta} \phi)^{\ast } \frac{ i \overleftrightarrow{\partial }^{\gamma } }{\sqrt{%
m^{2}-(\overrightarrow{\bigtriangleup })}+\sqrt{m^{2}-(\overleftarrow{%
\bigtriangleup })}} \phi . \label{eqtransss}
\end{eqnarray}
Finally, substituting the expression given in (\ref{cal-D}) into the above equation and combining the result with Eq.~(\ref{MM0}), we obtain (\ref{MPD11}).

\section{Vector current from the minimal substitution} \label{appenb}
\renewcommand{\theequation}{C.\arabic{equation}}
\setcounter{equation}{0}
$\;$ \vspace{-12pt}

The minimal substitution provides a gauge invariance of theory. After the minimal substitution the Lagrangian takes the form
\begin{equation}
{\rm {\mathcal L}}=\frac{1}{2} \phi ^{*} \left(
i\partial _{t} - A^{0} - \sum _{l=0}^{\infty } (-)^{l} f_{l} (m)(\mathbf{p}-\mathbf{A})^{2l} \right)\phi +{\rm c.c.},
\end{equation}
Based on the equation (\ref{off-ms}), we can immediately write ${\mathfrak{J}}^{0} = \phi ^{*} \phi$.


The variation of $S$ under variation of $\mathbf{A}$ can be found using the arguments similar to those
of Appendix~\ref{appen2}:
\begin{eqnarray}
\delta S &=&-\int d^{4}x\sum_{l=0}^{\infty
}\sum_{k=1}^{l}(-)^{l}f_{l}(m)\phi ^{\ast }(\mathbf{p})^{2k-2}(-p^{\alpha
}\delta A^{\alpha }\mathbf{-}\delta A^{\alpha }p^{\alpha })(\mathbf{p}%
)^{2l-2k}\phi  \nonumber \\
&=&\int d^{4}x\sum_{l=0}^{\infty }\sum_{k=1}^{l}f_{l}(m)\phi ^{\ast }(\overrightarrow{\Delta})
^{k-1}(i\overrightarrow{\partial }_{\alpha }\delta A^{\alpha }\mathbf{+}%
\delta A^{\alpha }i\overrightarrow{\partial }_{\alpha })(\overrightarrow{\Delta}) ^{l-k}\phi  \nonumber \\
&=&\int d^{4}x\delta A^{\alpha } \left( \sum_{l=0}^{\infty
}\sum_{k=1}^{l}f_{l}(m)\phi ^{\ast }(\overleftarrow{\Delta })^{k-1}i%
\overleftrightarrow{\partial }_{\alpha }(\overrightarrow{\Delta })^{l-k}\phi \right) \nonumber \\
&=&-\int d^{4}x\delta A^{\alpha }\left( \phi ^{\ast }i {\mathcal{D}}_{\alpha }\phi \right),
\end{eqnarray}
where we integrated by parts to remove derivatives from $\delta \mathbf{A}$. The bottom line is obtained using Eqs.~(\ref{cal-D}) and (\ref{binom}).

The current (\ref{off-ms}) coincides therefore with the Noether current (\ref{cov-D}).


\end{appendices}

\end{document}